\documentclass[12pt]{article}
\usepackage[latin1]{inputenc}
\usepackage[dvips]{graphicx}
\usepackage{graphicx}
\newcommand{\exprm}{{\rm exp}}

\newcommand{\drm}{{\rm d}}

\newcommand{\pa}{\partial}

\newcommand{\text}{\rm}

\newcommand{\ug}{ \; = \; }

\newcommand{\bb}{\begin{equation}}
\newcommand{\ee}{\end{equation}}
\newcommand{\bega}{\begin{eqnarray}}
\newcommand{\ega}{\end{eqnarray}}
\newcommand{\begae}{\begin{eqnarray*}}
\newcommand{\egae}{\end{eqnarray*}}

\newcommand{\h}{\hspace*{4ex}}
\newcommand{\dis}{\displaystyle}

\newcommand{\be}{\beta}

\newcommand{\om}{\omega}

\newcommand{\cent}{\centerline}
\newcommand{\vs}{\vspace*}

\newcommand{\brm}{\beta_{R_{m}}}
\newcommand{\bim}{\beta_{I_{m}}}

\newcommand{\krr}{k_{\rho R}}
\newcommand{\krm}{k_{\rho m}}
\newcommand{\krrm}{k_{\rho R_{m}}}
\newcommand{\krim}{k_{\rho I_{m}}}
\newcommand{\bm}{\beta_m}

\begin{document}

\baselineskip 0.8cm

\begin{center}

{\large {\bf Diffraction-Attenuation Resistant Beams: their Higher Order Versions and Finite-Aperture
Generations}$^{\: (\dag)}$} \footnotetext{$^{\: (\dag)}$ Work supported by CNPq and FAPESP. \ E-mail addresses
for contacts: mzamboni@ufabc.edu.br}


\end{center}

\vs{5mm}

\cent{ M. Zamboni-Rached, }

\vs{0.2 cm}

\centerline{{\em DMO--FEEC, State University at Campinas, Campinas, SP, Brazil.}}

\vs{0.5 cm}

\cent{ L. A. Ambrosio}

\vs{0.2 cm}

\centerline{{\em DMO--FEEC, State University at Campinas, Campinas, SP, Brazil.}}

\vs{0.2 cm}

\centerline{\rm and}

\vs{0.3 cm}

\cent{ H. E. Hern\'{a}ndez-Figueroa }

\vs{0.2 cm}

\cent{{\em DMO--FEEC, State University at Campinas, Campinas, SP, Brazil.}}

\vs{0.5 cm}

{\bf Abstract  \ --} \ Recently, a method for obtaining diffraction-attenuation resistant beams in absorbing
media was developed through suitable superposition of ideal zero-order Bessel beams. In this work, we will show
that such beams maintain their resistance to diffraction and absorption even when generated by finite apertures.
Also, we shall extend the original method to allow a higher control over the transverse intensity profile of the
beams. Although the method has been developed for scalar fields, it can be applied to paraxial vector wave
fields as well. These new beams can possess potential applications, such as free space optics, medical
apparatuses, remote sensing, optical tweezers, etc..





\section{Introduction}

About six years ago\cite{FW1,FW2}, an interesting method was developed, capable of delivering beams (in
non-absorbing media) whose longitudinal intensity pattern (LIP) could be previously chosen. This method, named
"Frozen Waves", is based on the superposition of co-propagating Bessel beams, all of them with the same
frequency.

\h A few time later, this method was further generalized\cite{FW3}, allowing us to model the LIP of propagating
beams in \emph{absorbing media}. As a particular case, diffraction-attenuation resistant beams were obtained,
that is, beams capable of maintaining both, the size and the intensity of their central spots for long distances
compared to ordinary beams.

\h The referred method for absorbing media was developed from appropriate superpositions of ideal zero-order
Bessel beams. This has two fundamental implications: a) beams with an infinite power flux (due to the use of
ideal Bessel beams), and b) beams with a spot-like transversal profile (due to the use of zero-order Bessel
beams).

\h In this paper, we shall extend the above method in such a way as to obtain a more efficient control over the
transversal intensity profile of such beams by adopting superposition of higher order Bessel beams.

\h We will also show that diffraction-attenuation resistant beams can maintain their interesting characteristics
even when generated by finite apertures, i.e., even when we transversally truncate the Bessel beams that
composes the desired beams. This demonstrates that, once the generation scheme is chosen, be it with antennas
(in microwaves and millimetric waves), holograms or spatial light modulators (in optics), the resultant beam may
possess characteristics analogous to those of the ideal case, at least until a certain field depth.

\h Although the method has been developed for scalar fields, it can be applied to electromagnetic waves in the
paraxial regime and we will elucidate this point.

\h These new beam solutions can possess potential applications in medicine, remote sensing, free space optics,
optical tweezers, etc..

\section{Resume of the method: Ideal Diffraction-Attenuation resistant beams in absorbing media}

In\cite{FW3} one of the authors developed a method capable of furnishing, \emph{in absorbing media}, beams that
are resistant to the effects of the diffraction and, the most important, capable of assuming any previously
chosen LIP, in $\rho=0$, in the range $0<z<L$.

\h As a particular case, this method is capable of furnishing diffraction-attenuation resistant beams for long
distances compared to ordinary beams. With "diffraction-attenuation resistant" we mean that the central spots of
these new beams maintain their shapes and also their intensities while propagating along an absorbing medium.

\h Such a method is based on suitable superpositions of equal-frequency zero-order Bessel beams and the core
idea is to take on extreme the self-reconstructing property of the non-diffracting waves
\cite{livro}-\cite{mrh2}. The obtained beams possess (initial) transversal field distributions capable of
reconstruct not only the shapes of the central spots, but also the intensities of these spots. This happens
without active action of the medium, which continues absorbing energy in the same way. This section presents the
method developed in\cite{FW3} with more details and some news.

\h In an absorbing media with a complex refractive index given by

\bb n(\om) \ug n_R(\om) + i n_I(\om) \ee

a zero-order Bessel beam can be written as

\bb \psi = J_0\left[(k_{\rho R}+k_{\rho I})\rho\right]{\rm exp}(i\be_R z){\rm exp}(-\be_I z){\rm exp}(-i\om t)
\label{bb}\ee

where $\be_R + i\be_I \equiv \be $ and $k_{\rho R}+k_{\rho I} \equiv k_{\rho} $ are the (complex) longitudinal
and transversal wave numbers, respectively, with their real and imaginary parts being given by:

\bb \be_R = \frac{n_R\om}{c} \cos\theta \,\, ;\,\,\, \be_I = \frac{n_I\om}{c} \cos\theta  \ee

and

\bb k_{\rho R} = \frac{n_R\om}{c} \sin\theta \,\, ;\,\,\, k_{\rho I} = \frac{n_I\om}{c} \sin\theta \,\,\, ,  \ee

being $0\leq \theta \leq \pi/2$ the axicon angle of the beam. Notice that $k_{\rho}^2 = n^2\om^2/c^2 - \beta^2$.

\h One can clearly see that the Bessel beam (\ref{bb}) suffers an exponential decay along the propagation
direction ``$z$'', due to the term ${\rm exp}(-\be_I z)$.

\h The absorption coefficient of a Bessel beam with an axicon angle $\theta$ is given by
$\alpha_{\theta}=2\beta_I=2n_I\om \cos\theta/c$, its penetration depth being
$\delta_{\theta}=1/\alpha_{\theta}=c/(2\om n_I\cos\theta)$. It is interesting to notice that, because the
transverse wave number $k_{\rho}$ is complex, the beam transverse profile decays as a Bessel function until
$\rho \approx 1/2k_{\rho I}$, beyond which it will suffer an exponential growth. This physically undesirable
behavior occurs because Eq.(\ref{bb}) represents an ideal Bessel beam which needs to be generated by an infinite
aperture. This problem, however, is solved when the beam is transversally truncated, i.e., when we use finite
apertures to its generation. In these cases, the exponential growth along the transverse direction (for
$\rho>1/2k_{\rho I}$) must cease for a given value of $\rho$, and when the radius $R$ of this aperture is such
that $R < 1/2k_{\rho I}$, this exponential growth does not even occurs\cite{FW3}.

\h It is important to remember that the efficient generation of a Bessel beam\cite{shep1,du} occurs when the
aperture radius\footnote{When generated by a finite aperture of radius $R>> 2.4/k_{\rho R}$ situated on the
plane $z=0$, the solution (\ref{bb}) becomes a valid approximation only in the spatial region $0 < z <
R/\tan\theta\equiv Z $ and to $\rho<(1-z/Z)R$} is such that $R >> 2.4/k_{\rho R}$.

\h From the two conditions for $R$ mentioned above, one can show\cite{FW3} that, in an absorbing medium, a
Bessel beam generated by a finite aperture of radius $R$ will possess acceptable characteristics when $n_R >>
n_I$, i.e., when the coefficient of absorption is such that $ \alpha << 1/\lambda \rightarrow \delta
>>\lambda $.

\h \emph{All cases considered here must obey this condition}.

\h Now that the basic characteristics of a Bessel beam in an absorbing medium are understood, let us present the
method developed in\cite{FW3}.

\h The idea is to achieve, in an absorbing medium with a refractive index $n(\om) \ug n_R(\om) + in_I(\om)$ , an
axially symmetric beam\footnote{In this paper we use cylindrical coordinates ($\rho,\phi,z$).},
$\Psi(\rho,z,t)$, whose LIP along the propagating axis (i.e., on $\rho=0$) can be freely chosen in a range
$0\leq z \leq L$. Let us say that the desired intensity profile in this range is given by $|F(z)|^2$. In order
to obtain a beam with such characteristics, the following solution is proposed:

\bb
\begin{array}{clr}
\Psi(\rho,z,t) & \ug \dis{\sum_{m=-N}^{N} A_m\,J_0\left(k_{\rho_m}\rho\right)\,e^{i\,\be_m z}\,e^{-i\,\om\,t}}\\
\\
& = \dis{e^{-i\,\om\,t}\sum_{m=-N}^{N} A_m\,J_0\left((k_{\rho R_m} + ik_{\rho
I_m})\rho\right)\,e^{i\,\be_{R_m}z}\,e^{-\be_{I_m}z}} \; ,
\end{array} \label{soma1} \ee

with

\bb
 k_{\rho_m}^2 \ug n^2\frac{\om^2}{c^2} - \be_{m}^2 \,\, ,\label{kr}
\ee

being

\bb \frac{\be_{R_m}}{\be_{I_m}} \ug \frac{\krrm}{\krim} \ug \frac{n_R}{n_I} \,\, , \label{bei} \ee

where $\be_m = \be_{R_m} + i\be_{I_m}$ and $k_{\rho_m} = k_{\rho R_m} + ik_{\rho I_m}$.

\h Equation (\ref{soma1}) is a superposition of $2N+1$ co-propagating Bessel beams with the same angular
frequency $\om$. In (\ref{soma1}), the coefficients $A_m$, the longitudinal ($\beta_m$) and the transverse
($k_{\rho\,m}$) wave numbers are yet to be determined. The choice of these values is made such that the desired
result (i.e., $|\Psi(\rho=0,z,t)|^2=|F(z)|^2$ within $0 \leq z \leq L$) is obtained.

\h The following choice is made:

\bb \be_{R_m} \ug Q + \frac{2\pi m}{L} \label{br} \,\, , \label{brm} \ee

where $Q$ is a constant such that

\bb 0 \leq Q + \frac{2\pi m}{L} \leq n_R \frac{\om}{c} \,\, , \label{cond} \ee

for $-N \leq m \leq N$.

\h Condition (\ref{cond}) ensures forward propagation only, with no evanescent waves. In (\ref{brm}) $Q$ is a
constant value that can be freely chosen, as long as (\ref{cond}) be obeyed and it plays an important role in
determining the spot size of the resulting beam (lower values of $Q$ implies in narrower spots) as we will show
soon. Besides, $Q$ can be chosen such as to guarantee the paraxial regime in an electromagnetic beam. In this
case, $Q$ must possess a value close to $n_R\om/c$. We will see this in details in section 3.

\h With the choice (\ref{brm}), the superposition (\ref{soma1}) is written as:

\bb
\begin{array}{clr}
\Psi(\rho,z,t) = e^{-i\,\om\,t}\,e^{i\,Qz}\, \dis{\sum_{m=-N}^{N}} A_m\,J_0\left((k_{\rho R_m} + ik_{\rho
I_m})\rho\right) \,e^{i\,\frac{2\pi m}{L}z}\,e^{-\be_{I_m}z}  \; ,
\end{array} \label{soma2} \ee

where, by inserting (\ref{brm}) into (\ref{bei}),

\bb \be_{I_m} \ug  \left(Q + \frac{2\pi m}{L}\right)\frac{n_I}{n_R} \,\, , \label{bei2} \ee

and $k_{\rho_m}=\krrm + i\krim$ is obtained through Eq.(\ref{kr}).

\h The maxima and minima of the imaginary parts of the various $\beta_{I\,m}$ are given by $(\be_I)_{\rm
min}=(Q-2\pi N/L)n_I/n_R$ and $(\be_I)_{\rm max}=(Q+2\pi N/L)n_I/n_R$, and the central value (for $m=0$) being
given by $(\be_I)_{m=0} = Q\, n_I/n_R \equiv \overline{\be}_I$.

\h Now, let us consider the ratio

\bb \Delta = \frac{(\be_I)_{\rm max} - (\be_I)_{\rm min}}{{\overline{\be}_I}} = 4 \pi \frac{N}{LQ}
\label{delta}\ee

\h For $\Delta <<1$, there are no considerable numerical differences among the various $\beta_{I\,m}$ and we can
safely approximate them by $\be_{I_m} \approx \overline{\be}_I$, which implies that $\exprm (-\be_{I_m}z)
\approx \exprm (-\overline{\be}_I z)$. In these cases, the series in Eq.(\ref{soma1}) evaluated on $\rho=0$ can
be (approximately) considered a truncated Fourier series, multiplied by the function $\exprm (-\overline{\be}_I
z)$. Therefore, this series can be used to reproduce the desired longitudinal profile $|F(z)|^2$ (on $\rho=0$),
within $0 \leq z \leq L$, as long as we make

\bb A_m \ug \frac{1}{L}\,\int_{0}^{L} F(z)\,e^{\overline{\be}_I z}e^{-i\,\frac{2\pi m}{L}z}\,dz \label{am} \ee

\h Essentially, this is the method developed in \cite{FW3}.

\h It is interesting to note that countless beams with the same desired LIP, but with different values of the
parameter $Q$ can be constructed. The basic difference among them will be their spot radius ($\Delta\rho$),
which can be estimated as being

\bb \Delta\rho \approx \frac{2.4}{k_{\rho R_{m=0}}} \ug \frac{2.4}{\dis{\sqrt{\frac{n_R^2\,\om^2}{c^2} - Q^2}}}
\label{Dr}\ee

\h So, besides choosing the desired LIP of the beam in an absorbing medium, we can stipulate its spot size. In
section 4 we will show that a more efficient control over the beam transverse intensity pattern is obtained by
using higher order Bessel beams in superposition (\ref{soma1}).

\h In resume, we wish to obtain a propagating beam in an \emph{absorbing medium} possessing, inside the interval
$0 \leq z \leq L$, a previously chosen LIP (on $\rho=0$) given by $|F(z)|^2$. To achieve such a profile, we
write the desired beam as a superposition of $2N+1$ co-propagating Bessel beams, Eq.(\ref{soma1}), by suitably
choosing the longitudinal ($\be_m = \brm + i\bim$) and transverse ($k_{\rho_m}=\krrm + i\krim$) wave numbers,
and the coefficients $A_m$ according to Eqs.(\ref{brm},\ref{cond},\ref{bei2},\ref{kr},\ref{am}). We can also
stipulate the spot radius, $\Delta\rho$, of the beam through a suitable choice of the parameter $Q$ in
Eq.(\ref{Dr}).

\h The method demonstrates to be efficient in situations where\footnote{Fortunately, these conditions are
satisfied for a great number of situations}$\alpha<<2/\lambda$ and $\Delta=4 \pi N/LQ <<1$, allowing us to
obtain a great variety of beams with potentially interesting intensity profiles as, for example, beams capable
of maintaining not only the size of their central spots, but also the intensity of these spots until a certain
chosen distance along the absorbing medium in question. We can call these types of beams as
``diffraction-attenuation resistant beams''.

\h We finish this section with an example\footnote{The same that was given in \cite{FW3}.} of the above
mentioned beam.

\textbf{Example}

\h Consider an absorbing medium with a refractive index $n = 1.5 + i\,0.49\times 10^{-6}$ in $\lambda = 308 {\rm
nm}$ (i.e., $\om = 6.12\times 10^{15}$Hz). In this bulk, at this angular frequency, we have the following
behaviors for the following wave solutions: a) A plane wave possesses a penetration depth of
$\delta=1/\alpha=c/(2\om n_I)=5\,$cm ; b) A gaussian beam with initial spot of radius $5.6\,\mu$m, besides
suffering attenuation, is also affected by a strong diffraction, having a diffraction length of only $0.6\,$mm;
c)An ideal Bessel beam with central spot of radius $5.6\,\mu$m (which implies in an axicon angle of $0.0141$rad)
can maintain its spot size, but suffer attenuation, possessing a penetration depth of
$\delta_{\theta}=1/\alpha_{\theta}=c/(2\om n_I\cos\theta)=5\,$cm.

\h Now, we are going to use the method exposed in this section to obtain a beam with spot radius $\Delta\rho=
5.6\,\mu$m, capable of maintaining the size and the intensity of its central spot until a distance of $25\,$cm,
i.e., a penetration depth $5$ times greater than those of the Bessel beam and of the plane wave\footnote{In an
absorbing medium like this, at a distance of $25\,$cm these beams would have got their initial field-intensity
attenuated $148$ times.}, and a diffractionless distance hundred times greater than that of the gaussian beam.
We also demand that the spot intensity suffers a strong fall after the distance $z=25\,$cm.

\h This diffraction attenuation resistant beam can be obtained through solution (\ref{soma1})  by choosing the
desired LIP $|F(z)|^2$ (on $\rho=0$), within $0 \leq z \leq L$, according to

\bb
 F(z) \ug \left\{\begin{array}{clr}
&1 \;\;\; {\rm for}\;\;\; 0 \leq z \leq Z  \\

&0 \;\;\; \mbox{elsewhere} ,
\end{array} \right.  \label{Fz1}
 \ee

putting $Z=25\;$cm, with, for example, $L=33\;$cm. The stipulated spot radius, $\Delta\rho= 5.6\,\mu$m, is
obtained by putting $Q=0.9999n_R\om / c$ in Eq.(\ref{Dr}). This value of $Q$ is used for the $\be_{R_m}$ in
Eq.(\ref{brm}), and\footnote{According to (\ref{cond}), the maximum value allowed for $N$ is $158$ and we choose
to use $N=20$ just for simplicity. Of course, using higher values of $N$ we get better results.} we choose
$N=20$. Finally we use Eq.(\ref{am}) to find out the coefficients\footnote{The analytic calculation of these
coefficients is quite simple in this case and their values are not listed here, we just use them in
Eq.(\ref{soma2}).} $A_m$ of (\ref{soma2}), defining in this way the desired beam.

\h We can see that the resulting beam fits very well all the desired characteristics, as is shown in Fig.1,
being 1(a) the 3D field-intensity and 1(b) the orthogonal projection in {\bf logaritmic} scale.

\h The resulting beam possesses a spot radius of $5.6\,\mu$m and maintains the size and intensity of its central
core till the desired distance of $25\,$cm, suffering after that an abrupt intensity fall.

\h As we will see in Section 5, the truncated version of this beam will maintain these characteristics if the
radius $R$ of the finite aperture used for truncation obeys $R \geq 3.8\,$mm. This is already suggested by the
Figure 1.b of the ideal (i.e., not truncated) beam.

\begin{figure}[!h]
\begin{center}
 \scalebox{2.7}{\includegraphics{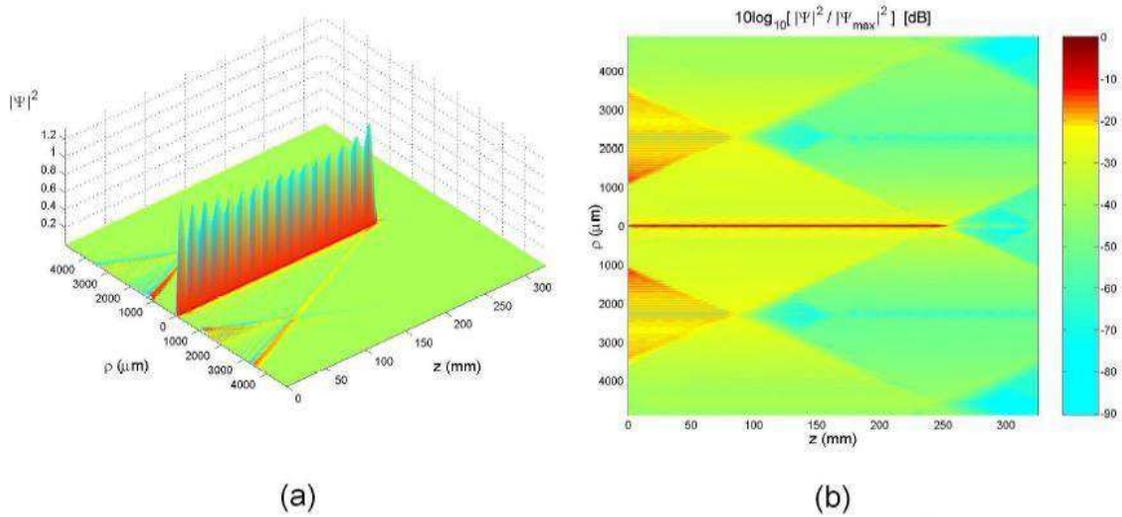}}
\end{center}
\caption{\textbf{(a)} 3D field-intensity of the resulting beam. \textbf{(b)} Orthogonal projection of the
resulting beam in {\bf logaritmic} scale.} \label{fig1}
\end{figure}


\

\section{Electromagnetic beams: The paraxial regime}

\h It should be clear to the reader that the present method is exact, i.e., the obtained beams (with the desired
LIPs) \emph{in absorbing media} are exact solutions to the scalar wave equation and can possess transverse spots
of any sizes, from wavelength dimensions to infinity.

\h In spite of the method has been developed to scalar fields, it can be used in electromagnetism (optics,
microwaves, etc..) \emph{in the paraxial regime}, where the scalar beam $\Psi$ of the previous section would
represent the transverse cartesian electric field component of a linearly polarized electromagnetic beam, being
$|\Psi|^2$ proportional to the time-averaged electromagnetic energy density. In these cases, the beam spot size
must be much greater than the correspondent wavelength. This can be done by choosing the parameter $Q$ of
Eq.(\ref{brm}) as $Q\approx n_R \om /c$.

\h We are going to elucidate all these points.

\h Consider an \emph{absorbing}, linear, homogeneous and isotropic medium without boundaries and without free
charges and free currents.

\h The electric (and magnetic) field obeys, in the monochromatic case, the Helmholtz equation:

\bb \nabla^2 \mathbf{E} + k^2\mathbf{E} \ug 0 \label{helm} \ee

with $\mathbf{E}= \mathcal{E}(x,y,z)e^{-i\om t}$ and where $k$ is the complex wave number:

\bb k \ug \om\sqrt{\mu(\om)\epsilon(\om)} \ug \om\sqrt{\mu\left(\epsilon_b + i\frac{\sigma}{\om}
 \right)} \ee

being $\epsilon_b(\om)$ the electric permittivity due to bound electrons, $\sigma(\om)$ the electric
conductivity and $\mu \approx \mu_0$ is the magnetic permeability.

\h Writing $k$ as

\bb k = k_R + ik_I \ee

and considering the frequency not so close to the resonance regions\footnote{In this case we can consider both
$\epsilon_b(\om)$ and $\sigma(\om)$ real quantities.}, we have\cite{jack}:

\bb k_R \approx \om \sqrt{\frac{\mu_0\epsilon_b}{2}}\left[\sqrt{1+\left(\frac{\sigma}{\epsilon_b\om} \right)^2
 }\,+1 \right]^{1/2} \ee

\bb k_I \approx \om \sqrt{\frac{\mu_0\epsilon_b}{2}}\left[\sqrt{1+\left(\frac{\sigma}{\epsilon_b\om} \right)^2
 }\,-1 \right]^{1/2} \ee

with the absorption coefficient\footnote{Notice that, according to section 2, the absorption coefficient of a
Bessel beam is $\alpha_{\theta}=\alpha\cos\theta=2k_I\cos\theta$. When $\theta \rightarrow 0$, the Bessel beam
tends to a plane wave and $\alpha_{\theta} \rightarrow \alpha$.} $\alpha=2k_I$.

\h Let $\mathbf{E}$ be an electric field given by:

\bb \mathbf{E} \ug E_x \,\mathbf{e}_x + E_z \,\mathbf{e}_z \,\, , \label{e} \ee

and let us apply our scalar method to the cartesian component $E_x$:


\bb \begin{array}{clr} E_x(\rho,z,t) & \ug \dis{e^{-i\,\om\,t}\,\sum_{m=-N}^{N}
A_m\,J_0(k_{\rho\,m}\,\rho)\,e^{i\be_m z}} \\

\\

& \ug \dis{e^{-i\,\om\,t}\,e^{i\,Q\,z}\,\sum_{m=-N}^{N} A_m\,J_0((\krrm +
i\krim)\,\rho)\,e^{-\be_{I_m}z}\,e^{i\,\frac{2\pi}{L}m\,z} }\,\, ,
\end{array}  \label{ex}
 \ee

\

where, as before, $\beta_m=\brm + i\bim$ and $\krm = \krrm + i\krim$ are chosen according to the
Eqs.(\ref{brm},\ref{cond},\ref{bei2},\ref{kr}), and $A_m$ calculated through Eq.(\ref{am}), being $|F(z)|^2$ the
desired LIP to the electric field component $E_x$.

\h Now, to satisfy the Gauss law $\nabla\cdot \mathbf{D}=0$ (no free charges), with $\mathbf{E} = E_x
\,\mathbf{e}_x + E_z \,\mathbf{e}_z$, in the monochromatic regime, we must have:

\bb E_z \ug -\int \frac{\pa E_x}{\pa x} dz  \ee

\h Using $E_x$ as giving by Eq.(\ref{ex}), we found:

\bb E_z(\rho,\phi,z,t) \ug \dis{e^{-i\,\om\,t}\,\sum_{m=-N}^{N}
A_m\,\frac{\krm}{\bm}\,J_1(k_{\rho\,m}\,\rho)\,\cos\phi\,e^{i\be_m z}} \label{ez} \ee

\h The paraxial regime is characterized by beams whose wave vectors of their constituent plane waves are almost
parallel to their propagation directions. In our case, this direction is ``$+z$'' and the paraxial regime is
reached by putting $Q\approx n_R \om/c$, implicating that $|\krm / \beta_m | << 1$, which in turn implicates
that $|E_z|<<|E_x|$. So, in this circumstance, we can use the so called paraxial approximation and write the
electric field (\ref{e}) as

\bb \mathbf{E} \approx E_x \,\mathbf{e}_x \ug \left(\dis{e^{-i\,\om\,t}\,\sum_{m=-N}^{N}
A_m\,J_0(k_{\rho\,m}\,\rho)\,e^{i\be_m z}}\right)\mathbf{e}_x \,\, , \label{epar} \ee

\hspace{10cm} (\emph{paraxial approximation})

\h The associated magnetic field can be found through Faraday's law:

\bb \mathbf{B} \ug -\,\frac{i}{\om}\,\nabla \times \mathbf{E}  \ee

\h Considering the electric field given by (\ref{e},\ref{ex},\ref{ez}), it is easy to show that in the paraxial
regime the above expression can be approximated as

\bb \mathbf{B} \, \approx \, \dis{-\frac{i}{\om} \frac{\pa E_x}{\pa z}} \mathbf{e}_y \ug
\left(\dis{e^{-i\,\om\,t}\,\sum_{m=-N}^{N} A_m\,\frac{\be_m}{\om}
  \,J_0(k_{\rho\,m}\,\rho)\,e^{i\be_m z}} \right)\mathbf{e}_y  \label{b}\ee

\hspace{10cm} (\emph{paraxial approximation})

\h Remembering that $\be_m=\brm + i \bim = \brm + i (n_I/n_R)\brm$, and that $0 \leq \brm \leq Q + 2\pi m/L$,
the paraxial regime ($Q\approx n_R \om /c$) implies that $\brm \approx n_R \om /c$, and in this case $\be_m/\om
\approx n/c\,$ (where $n=n_r + in_I$ is the complex refractive index). So, we can make another approximation in
Eq.(\ref{b}) writing:

\bb \mathbf{B} \, \approx \, \frac{n}{c}\left(\dis{e^{-i\,\om\,t}\,\sum_{m=-N}^{N} A_m
  \,J_0(k_{\rho\,m}\,\rho)\,e^{i\be_m z}} \right)\mathbf{e}_y  \ee

or

\bb \mathbf{B} \, \approx \, \frac{n}{c}\mathbf{e}_z \times \mathbf{E} \ug \frac{n}{c}E_x \mathbf{e}_y
\label{b2} \ee

\hspace{10cm} (\emph{paraxial approximation})

\h Using Eqs.(\ref{epar},\ref{b2}), we immediately see that the time-averaged energy density for monochromatic
electromagnetic fields, $u = (1/4) \rm{Re}(\mathbf{E}\cdot\mathbf{D}^* + \mathbf{B}\cdot\mathbf{H}^*)$, can be
approximated as

\bb u \, \approx \, \frac{1}{4}\rm{Re}\left(\epsilon^* + \frac{|n|^2}{\mu^*c^2} \right)|E_x|^2 \propto |E_x|^2
\label{u}\ee

\hspace{10cm} (\emph{paraxial approximation})

\h So, we can use our scalar method to obtain, in \emph{absorbing media}, paraxial electromagnetic beams whose
the time-averaged energy densities on the propagation axis can assume any desired patterns within an interval
$0\leq z \leq L$.

\

\section{Extending the method to nonaxially symmetric beams: Increasing the control over the transverse intensity pattern.}

\h The method developed in \cite{FW3} allows a strong control over the LIP (on $\rho=0$) of beams propagating in
\emph{absorbing media}.

\h Once we have controlled the beam's LIP, the transverse intensity pattern (TIP) can be shaped in a limited
way; more specifically, the spot size of the resultant beam can be chosen by a suitable choice of the parameter
$Q$ via Eq.(\ref{Dr}).

\h In this section\footnote[9]{The idea developed in this section generalizes that exposed in Section 5 of
reference \cite{FW2}, which was addressed to non-absorbing media.} we are going to show that it is possible to
get a more efficient control over the TIP, maintaining, at same time, a strong control over the LIP. It will be
possible, for instance, to shift the desired LIP from $\rho=0$ to $\rho=\rho'>0$. In other words, we will be
able to construct the desired LIP over cylindrical surfaces (instead of over the line $\rho=0$). Below we
explain this new procedure.

\h To obtain these new beams we proceed as before, choosing the desired LIP on $\rho=0$ within $0\leq z \leq L$,
choosing the values of $Q$ and $N$ (observing the Eq.(\ref{cond})), and calculating the values of $A_m$ through
Eq.(\ref{am}).

\h Having done this, we replace the zero order Bessel beams in superposition (\ref{soma2}) with higher order
ones. The new beam is written as:

\bb \dis{\Psi(\rho,\phi,z,t) \ug e^{-i\,\om\,t}\,e^{i\,Q\,z}\,e^{i\,\mu\,\phi}\,\sum_{m=-N}^{N}
A_m\,J_{\mu}(k_{\rho\,m}\,\rho)\,e^{-\be_{I_m}z}\,e^{i\,\frac{2\pi}{L}m\,z} } \; , \label{soma4} \ee

with $\mu$ a positive integer and all other parameters ($Q$, $N$, $L$ and $A_m$) given and calculated as before.

\h For the situations considered here, we have that $n_I<<n_R \rightarrow \krim<<\krrm $. This implies that for
each mth Bessel function in Eq.(\ref{soma2}) is valid: $J_{\mu}[(\krrm + i\krim)\rho] \approx J_{\mu}(\krrm\rho)
$ for $0 < \rho << 1/\krim$, and in this range each mth Bessel function reaches its maximum value at
$\rho=\rho'_m$, being $\rho'_m$ the first positive root of the equation
$(\drm\,J_{\mu}(\krrm\rho)/\drm\rho)|_{\rho'_m}=0$.

\h The values of $\rho'_m$ are located around the central value $\rho'_{m=0}$ and we can expect a shift of the
desired LIP from $\rho=0$ to $\rho\approx\rho'_{m=0}$.

\h We have confirmed that conjecture in all situations considered by us, mainly in the cases where there is no
considerable difference among the values of $\krrm$.

\h With this extension of the original method, it is possible to model the LIP over cylindrical surfaces,
obtaining very interesting static configuration of the field intensity. In particular, we can construct (in
absorbing media) hollow beams resistant to the attenuation and diffraction effects.

\h To show this method working we are going to obtain a cylindrical light surface of constant intensity in an
absorbing medium with $n_R = 1.5$ and $\alpha = 20 {\rm m}^{-1}$ (at $\lambda = 308$nm $\rightarrow$ $\om  =
6.12\times 10^{15}$Hz), which implies that $n_I = 0.49\times 10^{-6}$. At this angular frequency, an ordinary
beam would possess a depth of penetration of $5\,$cm in this medium.

\h Let us choose, within $0\leq z \leq L$, a LIP as that of section 1, Eq.(\ref{Fz1}):

\bb
 F(z) \ug \left\{\begin{array}{clr}
&1 \;\;\; {\rm for}\;\;\; 0 \leq z \leq Z  \\

&0 \;\;\; \mbox{elsewhere} ,
\end{array} \right.  \label{Fz2}
 \ee

being $Z=25\;$cm and $L=33\;$cm.

\h As in section 1, we put $Q=0.9999n_R \om /c$ in Eq.(\ref{brm}). With the chosen values to $Q$ and $L$, the
maximum value allowed to $N$ is $N=154$, but for simplicity reasons we choose $N=20$.

\h Using equations(\ref{brm},\ref{cond},\ref{bei2},\ref{kr},\ref{am}) we evaluate all the $\bm$, $\krm$ and
$A_m$. But, as we have explained in this section, instead of using all these values in (\ref{soma2}), we use
them in the superposition (\ref{soma4}), where we choose $\mu=4$.

\h According to the previous discussion, we can expect the desired LIP over a cylindrical surface of radius
$\rho \approx 5,318/k_{\rho R_{m=0}} = 12.289 \,\mu$m (that is where the function $J_4(k_{\rho R_{m=0}}\rho)$
assumes its maximum value).

\h We can see in figure(2.a) the resulting intensity field in a three-dimensional pattern. Its orthogonal
projection shown in Fig.(2.b) clearly confirms the cylindrical surface of light intensity. Figure (2.c) depicts
the transverse intensity pattern at $z=L/2$. It is possible to note that the transverse peak intensity is
located at $\rho = 12.285 \,\mu$m, a value near to the predicted one.

\begin{figure}[!h]
\begin{center}
 \scalebox{3.3}{\includegraphics{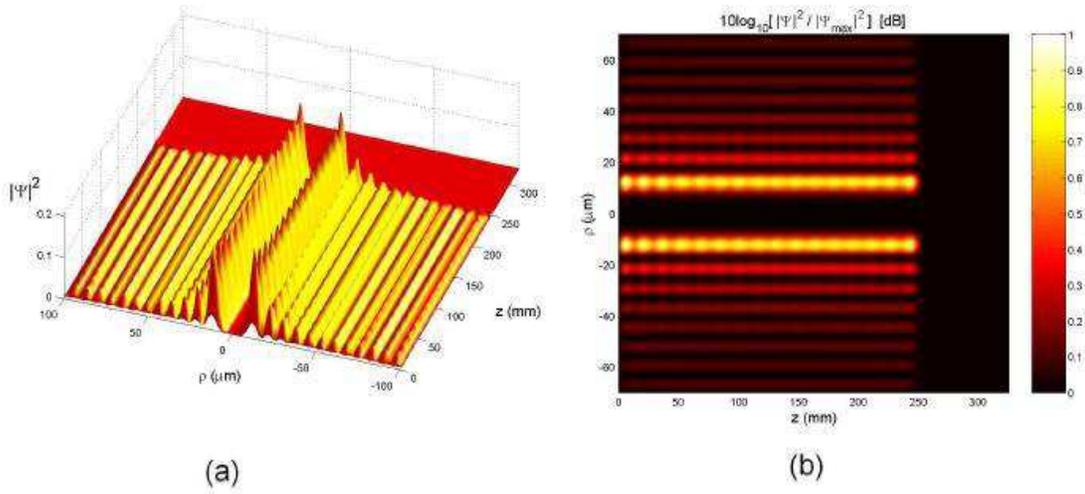}}
 \end{center}
\caption{\textbf{(a)} 3D field-intensity of the resulting beam. \textbf{(b)} Its orthogonal projection.}
\label{fig2}
\end{figure}

\

\begin{figure}[!h]
\begin{center}
 \scalebox{1.2}{\includegraphics{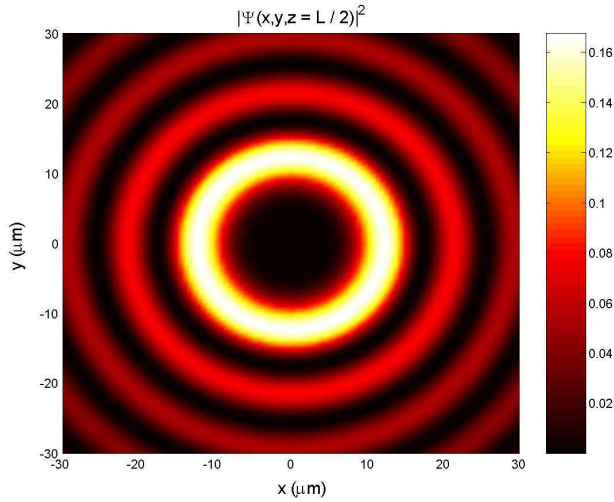}}
 \end{center}
\caption{The beam's transverse intensity pattern at $z=L/2$} \label{fig3}
\end{figure}

\h This interesting field configuration is resistant to the attenuation and diffraction effects till the
distance $z=25$cm. We should remember that any other ordinary beam at the same frequency, propagating in the
same medium, would have a penetration depth of only $5$cm.

\

\section{Finite aperture generation of the diffraction-attenuation resistant beams}

\h As we have seen, the solution (\ref{soma1}), with (\ref{brm},\ref{cond},\ref{bei2},\ref{kr},\ref{am}),
represents propagating beams in absorbing media with the remarkable characteristic of allowing us to choose the
desired LIP on $\rho=0$, within $0 \leq z \leq L$, being the spot sizes regulated by the value of the parameter
$Q$. The same occurs with solution (\ref{soma4}), which in turn allows us to choose the desired LIP on a
cylindrical surface.

\h Now, we must remember that, in spite of the beams (\ref{soma1}) and (\ref{soma4}) be exact solutions, they do
not represent beams generated or truncated by finite apertures. Actually, fields given by those solutions
\emph{in all points of space} would require infinite apertures to be generated.

\h However, if a Bessel beam given by Eq.(\ref{bb}) is truncated by a finite aperture of radius $R>>2.4/\krr$
situated on the plane $z=0$, the radiated field -- in the spatial region $0 < z < R/\tan\theta \equiv Z$ (the
diffractionless distance of a truncated Bessel beam) and $0 \leq \rho \leq (1-z/Z)R\,$ -- can be approximately
described\footnote[8]{The same is valid for a truncated higher-order Bessel beam.}\cite{FW2,FW3,du} by
Eq.(\ref{bb}).

\h Taking into account that the solution (\ref{soma2}) is a linear superposition of Bessel beams, we can expect
that when it is truncated by a finite aperture of radius\footnote[5]{Notice that $k_{\rho R_{m=N}}$ is the
smallest value of all $\krrm$, therefore if $R>>2.4/k_{\rho R_{m=N}} \rightarrow R>>2.4/\krrm$ for all $m$.}
$R>>2.4/k_{\rho R_{m=N}}$, the radiated field -- in the region\footnote[6]{Here, $\theta_m$ is the axicon angle
of the mth Bessel beam in (\ref{soma1}).} $0 < z < R/\tan\theta_{m=-N} \equiv Z_{m=-N}$ and $0 \leq \rho \leq
(1-z/Z_{m=-N})R\,$ -- will be approximately given by  Eq.(\ref{soma2}).

\h Now, as we are interested in controlling the LIP of the truncated beam within $0 \leq z \leq L$ , we have to
guarantee that, after the truncation, all Bessel beams in (\ref{soma2}) maintain their characteristics until
$z=L$. This is possible if both conditions are satisfied:

\bb  R >> \frac{2.4}{k_{\rho R_{m=N}}} \label{R1}  \ee

and\footnote[7]{That is, the shortest diffractionless distance is larger than the distance $L$.}

\bb Z_{m=-N} > L \,\,  \rightarrow \,\, \frac{R}{\tan\theta_{m=-N}}>L \,\, \rightarrow \,\, R >
L\sqrt{\frac{n_R^2}{c^2}\frac{\om^2}{\be_{R_{m=-N}}^2} -1 } \label{R2} \ee

\h So we can \emph{expect} that by choosing a finite aperture of radius $R$ large enough to satisfy (\ref {R1})
and (\ref{R2}), the truncated versions of the diffraction attenuation resistant beams will maintain their
characteristics, i.e., we will continue to be able to control the beam's LIP.

\h In order to confirm our expectation, some examples shall be presented, where we choose some desired LIPs and
obtain the correspondent ideal beam solutions, $\Psi(\rho,\phi,z,t)$, through
Eqs.(\ref{soma1},\ref{brm},\ref{cond},\ref{bei2},\ref{kr},\ref{am}). These ideal solutions, in turn, are used
for obtaining their truncated versions, $\Psi_T(\rho,\phi,z,t)$, through numerical calculation of the
Rayleigh-Sommerfeld diffraction integral for monochromatic waves\cite{Goodman}:

\bb \Psi_T(\rho,\phi,z,t) \ug  \frac{1}{2\pi}
\int_0^{2\pi}d\phi'\int_0^{R}d\rho'\,\rho'\,\frac{e^{ikD}}{D}\left(\frac{\pa }{\pa
z'}\,\Psi(\rho',\phi',z',t)\right)_{z'=0} \,\,\, , \label{int}   \ee

where a circular aperture of radius $R$, on the plane $z'=0$, is used for the truncation, being the distance
$D=\sqrt{\left(z-z'\right)^2+\rho^2+\rho'^2-2\rho\rho'\cos\left(\phi-\phi'%
\right)}$ the separation between the source and observation points.

\

\textbf{A. First Example}

\h Here we are going to calculate numerically the truncated version of the ideal diffraction-attenuation
resistant beam obtained in the example of Section II and in ref.\cite{FW3}.

\h The medium in question has refractive index $n = n_R + in_I=1.5 + i\,0.49\times 10^{-6}$ in $\lambda = 308
{\rm nm}$ (i.e., $\om = 6.12\times 10^{15}$Hz). The \emph{ideal} beam in that example was constructed to possess
(in $\om = 6.12\times 10^{15}$Hz) a spot radius $\Delta\rho= 5.6\,\mu$m and a constant intensity of its central
spot until a distance of $25\,$cm. These characteristics were reached through the \emph{fundamental ideal
solution} (\ref{soma2}), with Eqs.(\ref{brm},\ref{cond},\ref{bei2},\ref{kr},\ref{am}), by choosing the desired
LIP $|F(z)|^2$ (on $\rho=0$), within $0 \leq z \leq L$, according to Eq.(\ref{Fz1}), putting $Z=25\;$cm and
$L=33\;$cm. The values of $Q$ and $N$ were chosen to be $0.9999n_R\om / c$ and $20$ respectively.

\h This ideal beam solution is used as the aperture excitation, $\Psi(\rho',z',t)$, in the Rayleigh-Sommerfeld
diffraction integral, Eq.(\ref{int}), which is numerically calculated to yield $\Psi_T(\rho,z,t)$, i.e., the
truncated version of this beam.

\h According to our above discussion, the truncated beam shall possesses a behavior very similar to the ideal
one provided that the aperture radius $R$ satisfy both conditions (\ref{R1}, \ref{R2}), which furnishes in this
case $R\geq 4.9\,$mm. However, due to the fact that the chosen on-axis LIP has null value within $Z < z < L$, we
can replace $L$ in condition (\ref{R2}) with $Z$, obtaining in this case $R \geq 3.8\,$mm. We choose $R =
3.8\,$mm.

\h After numerical calculation of (\ref{int}), we obtain the result plotted below. Comparing Figure(3) with
Figure(1a) of the ideal beam of Section II, we can see that our expectations were correct, i.e., by choosing an
aperture radius big enough, the truncated version becomes very close to the ideal solution in the spatial region
of interest.

\begin{figure}[!h]
\begin{center}
 \scalebox{1.8}{\includegraphics{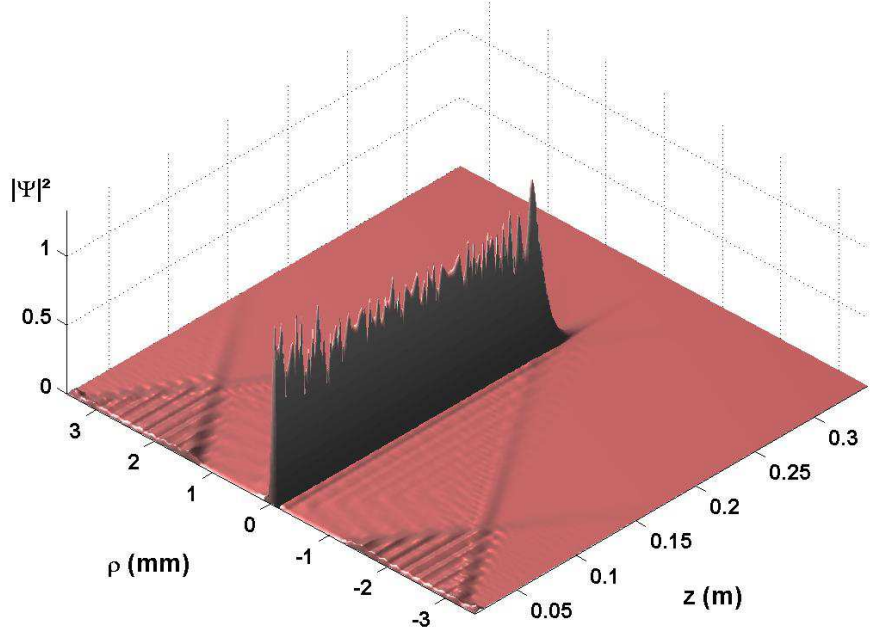}}
 \end{center}
\caption{The truncated version of the ideal diffraction-attenuation resistant beam obtained in the example of
Section II} \label{fig4}
\end{figure}

\

\textbf{B. Second Example}

\h In \cite{FW3} one of the authors obtained, in an absorbing media, an ideal (i.e., not truncated) beam
presenting an interesting and counterintuitive characteristic. There, it was considered a medium with refractive
index $n = n_R + in_I = 1.5 + i0.46 \times 10^{-6}$ (which implies in a penetration depth of $5\,$cm) at $\om =
6.12\times 10^{15}\,$Hz. At this angular frequency an ideal beam was shaped to possess a spot radius
$\Delta\rho= 5.6\,\mu$m and a modest \emph{exponential growth} of its intensity until a distance of $25\,$cm,
suffering after this a strong intensity fall.

\h In order to reach these characteristics, the desired on axis LIP, $|F(z)|^2$, within $0 \leq z \leq L$, was
chosen according to:

\bb
 F(z) \ug \left\{\begin{array}{clr}
& {\rm exp}(z/Z) \;\;\; {\rm for}\;\;\; 0 \leq z \leq Z  \\

&0 \;\;\; \mbox{elsewhere} ,
\end{array} \right.  \label{Fz3}
 \ee

with $Z=25\,$cm and $L=33\;$cm. Taking into account Eq.(\ref{Dr}), it was chosen $Q=0.9999n_R\om / c$ and the
value of $N$ was chosen to be $N=20$.

\h With this, the (ideal) desired beam, $\Psi(\rho,z,t)$, was obtained in \cite{FW3} through the fundamental
solution (\ref{soma2}), with Eqs.(\ref{brm},\ref{cond},\ref{bei2},\ref{kr},\ref{am}). Below we plotted this
resulting ideal (i.e., not truncated) beam.

\begin{figure}[!h]
\begin{center}
 \scalebox{1.3}{\includegraphics{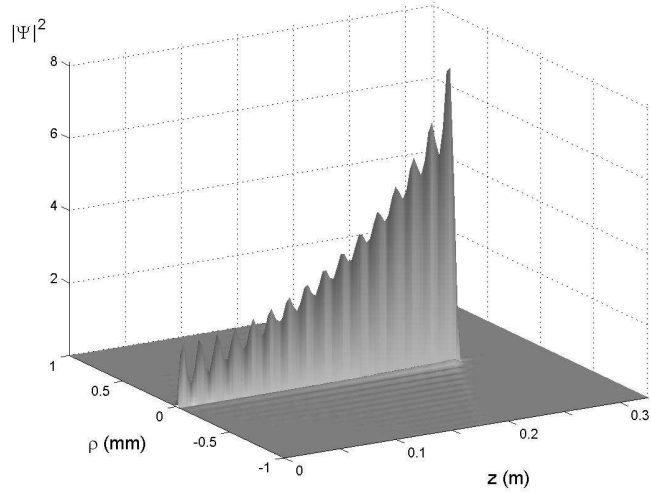}}
 \end{center}
\caption{Ideal beam presenting a moderate exponential \emph{growth} in an absorbing medium} \label{fig5}
\end{figure}

\h In this example, we are going to use Eq.(\ref{int}) to obtain the truncated version, $\Psi_T(\rho,z,t)$, of
the above ideal beam.

\h By using conditions (\ref{R1}, \ref{R2}) for an efficient finite aperture generation, we obtain that the
aperture radius should obey $R\geq 4.9\,$mm. But, for the same reason of the previous example, we can adopt
$R\geq 3.8\,$mm, and we choose $R=3.8\,$mm.

\h The numerical calculation of Eq.(\ref{int}) yields the truncated beam plotted below:

\begin{figure}[!h]
\begin{center}
 \scalebox{1.3}{\includegraphics{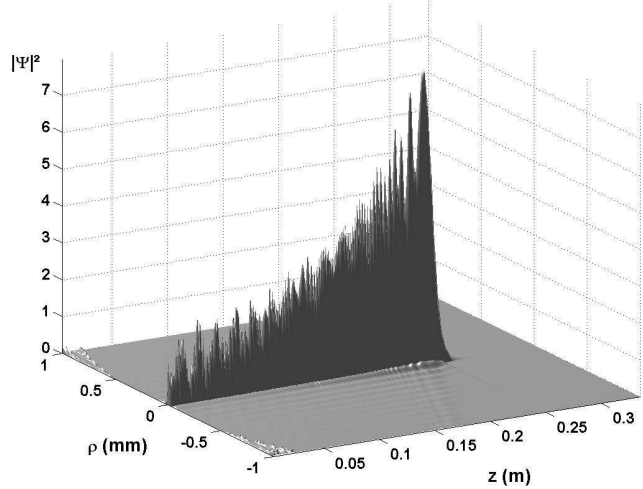}}
 \end{center}
\caption{Truncated version of the beam with exponential growth} \label{fig6}
\end{figure}

\h We can see an excellent agreement between the ideal beam and the truncated one in the region of interest,
confirming that the method works very well in more realistic situations close to the experimental ones.

\

\section{Conclusions}

\h A few years ago\cite{FW3} it was developed an interesting theoretical method, from which it is possible to
construct, in \emph{absorbing media}, axially symmetric beams whose the LIP can be previously chosen. As a
particular case, diffraction-attenuation resistant beams were obtained, that is, beams capable of maintaining
both, the size and the intensity of their central spots.

\h In this paper we elucidate how to apply this scalar method in paraxial electromagnetic waves and extended it
to include non-axially symmetric beams, allowing in this way, besides the strong control over the longitudinal
intensity patter, a certain control over the transverse one. With this extension, it is possible to model the
LIP over cylindrical surfaces. In particular, we can construct (in absorbing media) hollow beams resistant to
the attenuation and diffraction effects.

\h We also use the Rayleigh-Sommerfeld diffraction integral to obtain the truncated versions of these new beams,
verifying that, provided that the aperture used for truncation is big enough, the truncated beams possess the
same interesting characteristics of the ideal beams. This verification is important because it confirms that the
method works very well in more realistic situations close to the experimental ones.

\h These new beams can possess potential applications, such as free space optics, medical apparatuses, remote
sensing, etc..

\

\section*{Acknowledgements}

The authors are very grateful, for collaboration and many stimulating discussions over the last few years, to
Erasmo Recami. This work has been partially supported by CNPq and FAPESP (Brazil).

\

\

\end{document}